\documentclass[paper]{aa} % for the letters 
\usepackage{graphicx}
\usepackage{txfonts}
\usepackage{longtable}
\usepackage{natbib}
\usepackage{url}
\bibpunct{(}{)}{,}{a}{}{,}
\begin{document}
\title{{Heavy water stratification in a low-mass protostar} \thanks{Based on \textit{Herschel}/HIFI observations. {\it Herschel}
    is an ESA space observatory with scientific instruments provided by
    European-led principal Investigator consortia and with important
    participation from NASA.}}

  % \subtitle{}

   \author{
A. Coutens \inst{1,2} \and
C. Vastel \inst{1,2} \and S. Cazaux \inst{3} \and S. Bottinelli \inst{1,2} \and E. Caux \inst{1,2} \and C. Ceccarelli \inst{4} \and K. Demyk \inst{1,2} \and V. Taquet  \inst{4} \and V. Wakelam \inst{5}
}

   \institute{Universit\'e de Toulouse, UPS-OMP, IRAP, Toulouse, France    \\
    \email{acoutens@nbi.dk, cvastel@irap.omp.eu} 
         \and
             CNRS, Institut de Recherche en Astrophysique et Plan\'etologie, 9 Av. Colonel Roche, BP 44346, 31028 Toulouse Cedex 4, France 
             \and
            Kapteyn Astronomical Institute, P.O. Box 800, 9700AV Groningen, The Netherlands  
             \and Institut de Plan\'etologie et d'Astrophysique de Grenoble, UMR 5274, UJF-Grenoble 1/CNRS, 38041 Grenoble, France      
             \and CNRS and Universit\'e de Bordeaux, Observatoire Aquitain des Sciences de l'Univers, 2 rue de l'Observatoire, B.P. 89, F-33271 Floirac, France
}

   \date{Received 20 December 2012; accepted 19 March 2013}

  \abstract
  {Despite the low elemental deuterium abundance in the Galaxy,
    enhanced molecular deuterium fractionation has been found in the environments
    of low-mass star-forming regions and, in particular, the Class 0
    protostar IRAS~16293-2422.}
  {The key program Chemical HErschel Surveys of Star forming regions (CHESS) aims at studying the molecular complexity of the
    interstellar medium. The high sensitivity and spectral resolution
    of the {\it Herschel}/HIFI (Heterodyne Instrument for Far-Infrared) instrument provide a unique opportunity to observe the
    fundamental 1$_{1,1}$--0$_{0,0}$ transition of ortho--D$_2$O at 607 GHz
  and the higher energy 2$_{1,2}$--1$_{0,1}$ transition 
   of para--D$_2$O  at 898 GHz, both of which are inaccessible from the ground.
}
 {The ortho--D$_2$O transition at 607 GHz was previously detected. We present in this paper the first tentative detection for the para--D$_2$O transition at 898 GHz. 
 The spherical Monte Carlo radiative transfer code RATRAN was used to reproduce the observed line profiles of D$_2$O with the same method
 that was used to reproduce the HDO and H$_2$$^{18}$O line profiles in IRAS~16293-2422. 
 }
 { As for HDO, the absorption component seen on the D$_2$O lines can only be reproduced by adding an external absorbing layer, possibly created by the photodesorption 
 of the ices at the edges of the molecular cloud. The D$_2$O column density is found to be about 2.5 $\times$ 10$^{12}$ cm$^{-2}$ in this added layer, leading to a D$_2$O/H$_2$O ratio of about 0.5\%. At a 3$\sigma$ uncertainty, upper limits of 0.03\% and 0.2\% are obtained for this ratio in the hot corino and the colder envelope of IRAS~16293-2422, respectively.  }
    {The deuterium fractionation derived in our study suggests that the ices present in IRAS~16293-2422 formed on warm dust grains ($\sim$ 15 -- 20 K) in dense ($\sim$ 10$^4$ -- 5 $\times$ 10$^4$  cm$^{-3}$)  translucent clouds. These results allow us to address the earliest phases of star formation and the conditions in which ices form. }


   \maketitle
%
%________________________________________________________________

\section{Introduction}

Despite the low elemental deuterium abundance in the local interstellar gas \citep[1--2 $\times$ 10$^{-5}$;][]{Linsky2006}, numerous deuterated molecules have been detected with enhanced D/H ratios in the environments of low-mass star-forming regions. The doubly deuterated form of water, D$_2$O, has been in particular detected in the solar-type protostar IRAS~16293-2422 (hereafter IRAS~16293), which is well known for its high deuterium fractionation: the fundamental 1$_{1,0}$--1$_{0,1}$ para--D$_2$O transition at 317 GHz was detected by \citet{butner2007} 
using the James Clerk Maxwell Telescope (JCMT), whereas the fundamental 1$_{1,1}$--0$_{0,0}$ 
ortho--D$_2$O transition at 607 GHz was discovered with the Heterodyne Instrument for Far-Infrared (HIFI) instrument onboard the \textit{Herschel}
Space Observatory by \citet{vastel2010}. 
The observed line profile at 317 GHz shows a component in emission 
in addition to a deep and narrow absorption. The emission component has 
been attributed to heavy water in the hot corino of this source, where the grain ices are sublimated 
and released into the gas phase \citep{ceccarelli2000,bottinelli2004}. The absorption component, 
whose linewidth is 0.5 km\,s$^{-1}$, is seen in both transitions at the velocity of 4.2 km\,s$^{-1}$. 
It probably originates in the foreground gas (molecular cloud and cold envelope), which is studied through 
the fundamental HDO transition at 465 and 894 GHz \citep{coutens2012}. Both transitions present 
deep absorption features at the same velocity as the D$_2$O component. The OD molecule, a key species in the formation of deuterated water on grains, also shows a deep absorbing component \citep{Parise2012} and most certainly arises from the same region as water and its deuterated forms. 
Besides, numerous absorption lines of several nitrogen molecules (NH, ND, NH$_2$, and NH$_3$) are detected towards the IRAS~16293 source \citep{bacmann2010,Hily-Blant2010}, suggesting an origin of these species in a cold environment. 
\citet{coutens2012} used the
spherical Monte Carlo radiative transfer code RATRAN \citep{ratran} to reproduce the line 
profiles of the two HDO fundamental transitions as well as eleven other detected lines by assuming an abundance jump at 100 K \citep{Fraser2001}. 
From the emission line profiles, the derived HDO inner abundance (T $\geq$ 100 K) is about 1.7 $\times$ 10$^{-7}$, whereas the HDO outer abundance (T $<$100 K) is estimated at about 8 $\times$ 10$^{-11}$.
A water-rich absorbing layer is, however, required in front of the envelope to reproduce the HDO absorption components observed at 465 and 894 GHz. This layer may result from the photodesorption of the ices at the edges of the molecular cloud, as predicted by chemical models \citep[e.g.,][]{Hollenbach2009}.

A non-LTE (local thermal equilibrium) modeling has been performed by \citet{vastel2010} to estimate the ortho-to-para 
D$_2$O ratio in the cold ($<$ 30 K) dense ($<$ 5 $\times$ 10$^6$ cm$^{-3}$) cloud surrounding the 
IRAS~16293 protostar, adopting the density and temperature profiles of the envelope \citep{crimier2010}. 
\citet{vastel2010} considered the first two levels of each D$_2$O form and used the computed collisional rates 
for the two fundamental deexcitation transitions of ortho-- and para--D$_2$O with para--H$_2$ in the 10--30 K 
range \citep{Scribano2010}.
Their computation yields an ortho-to-para ratio of 1.1 $\pm$ 0.4 at a 1$\sigma$ level of uncertainty 
with the corresponding column densities N(ortho) = (8.7 $\pm$ 2.1) $\times$ 10$^{11}$ cm$^{-2}$ and 
N(para) = (7.8 $\pm$ 2.6) $\times$ 10$^{11}$ cm$^{-2}$. This ratio is lower than 2.6 at a 3$\sigma$ level of 
uncertainty, taking into account the statistical error as well as the overall calibration budget for the HIFI 
band 1b. The comparison between the upper value of the measured D$_2$O ortho-to-para ratio and the 
thermal equilibrium value shows that they are consistent with a gas at a temperature higher than
about 15 K and, therefore, with the assumed absorbing gas location. 

The present paper aims at determining the D$_2$O abundance distribution throughout the whole protostellar envelope of IRAS~16293, from the hot corino to the foreground water-rich absorbing layer (see Figure 2 in \citealt{coutens2012}).
Using the previous results obtained in a similar way by \citet{coutens2012} for HDO and H$_2$O, the derived D$_2$O/H$_2$O and HDO/H$_2$O ratios are then discussed and compared with a grain-surface chemical model \citep{cazaux2011}.
The paper is organized as follows. First, we describe the observations as well as their reduction in Section 2. Then we present the modeling and the results in Section 3. Finally, we discuss the results in Section 4 and conclude in Section 5.

\section{Observations}
\label{sect_obs}

In the framework of the Chemical HErschel Surveys of Star forming regions (CHESS) key program \citep{ceccarelli2010}, we observed the solar-type protostar IRAS~16293 with the \textit{Herschel}/HIFI
instrument \citep{pilbratt2010,degraauw2010}. The targeted coordinates were $\alpha_{2000}$ = 16$^h$ 32$^m$ 22$\fs$75, $\delta_{2000}$ = $-$ 24$\degr$ 28$\arcmin$ 34.2$\arcsec$, a position at equal distance from the binary components A and B.
A full spectral coverage of bands 1b and 3b was performed on 2010 March 2 and 2010 March 19 respectively, using the HIFI spectral scan double beam switch (DBS) mode with optimization of the continuum. The fundamental ortho--D$_2$O 1$_{1,1}$--0$_{0,0}$ transition lies in this frequency
coverage at 607.349 GHz  whereas the para--D$_2$O 2$_{1,2}$--1$_{1,0}$ transition lies at 897.947 GHz. The HIFI Wide
Band Spectrometer (WBS) was used, providing a spectral resolution of 1.1 MHz ($\sim$0.55 km\,s$^{-1}$ at 600 GHz and $\sim$0.37 km\,s$^{-1}$ at 900 GHz) over an instantaneous
bandwidth of 4~$\times$~1~GHz. %The data are acquired at the Nyquist sampling, therefore, with 0.5 MHz steps. 
The ortho--D$_2$O 1$_{1,1}$--0$_{0,0}$ transition was re-observed in %4 February 2011
February 2011 with the High Resolution Spectrometer (HRS) to benefit from a higher spectral resolution ($\sim$0.06 km\,s$^{-1}$). We consequently use these new observations hereafter.
At 610 GHz (900 GHz), the beam size is about 35$\arcsec$ (24$\arcsec$ resp.) and the measured main beam efficiency is 0.75 (0.74 resp.), whereas the forward efficiency is 0.96 in both cases \citep{roelfsema2012}. The DBS reference positions were situated approximately 3$\arcmin$ east and west of the source. 

The data reduction of the para--D$_2$O 2$_{1,2}$--1$_{1,0}$ transition at 898 GHz was performed similarly to the HDO and H$_2$$^{18}$O lines studied by \citet{coutens2012}.
This transition was observed eight times (four in lower sideband and four in upper sideband) for each polarization in the selected observing mode. To produce the final spectra, the observations that were processed using the standard HIFI pipeline up to level 2 with the ESA-supported package Herschel Interactive Processing Environment (HIPE) 5.1 \citep{ott2010} were exported to the GILDAS/CLASS\footnote{\url{http://www.iram.fr/IRAMFR/GILDAS}} software. We then baseline subtracted and averaged the H and V polarizations at the line frequency, weighting them by the observed noise for each spectra. We verified for all spectra that no emission from other species was present in the image band. 
This transition had not been discovered when \citet{vastel2010} published 
their results with an early version (3.01) of the HIPE software, using the sideband deconvolution method 
on all bands. %Note that this method, from HIPE 3.01 to HIPE 8.0, leads to the same non-detection result and noise level. 
The present data reduction, which uses the selected spectra re-aligned on one single transition and averaged after an accurate baseline subtraction and rms weighting, leads to a lower noise level 
than the one obtained with the standard HIPE deconvolution routine, and allows us to present the first tentative detection of a high-energy level transition of deuterated water (see Fig. \ref{observations}).
For the pointed HRS data at 607 GHz, the observations of the H and V polarizations were processed with the ESA-supported package HIPE 8.0 \citep{ott2010} up to level 2 and then exported to the GILDAS/CLASS software for the baselines to be subtracted and the two polarizations averaged. 
HIFI operates as a double sideband (DSB) receiver. From the in-orbit performances of the instrument \citep{roelfsema2012}, a sideband ratio of unity is assumed for the observed D$_2$O transitions seen in absorption against the continuum (bands 1b and 3b). The continuum values observed over the frequency range of the whole bands are closely fitted by straight lines. 
The continuum derived from the polynomial fit at the considered frequency and divided by two to obtain the single sideband (SSB) continuum was added to the spectra.  
 
\begin{figure}[t!]
  \centering
  \includegraphics[width=8.5cm]{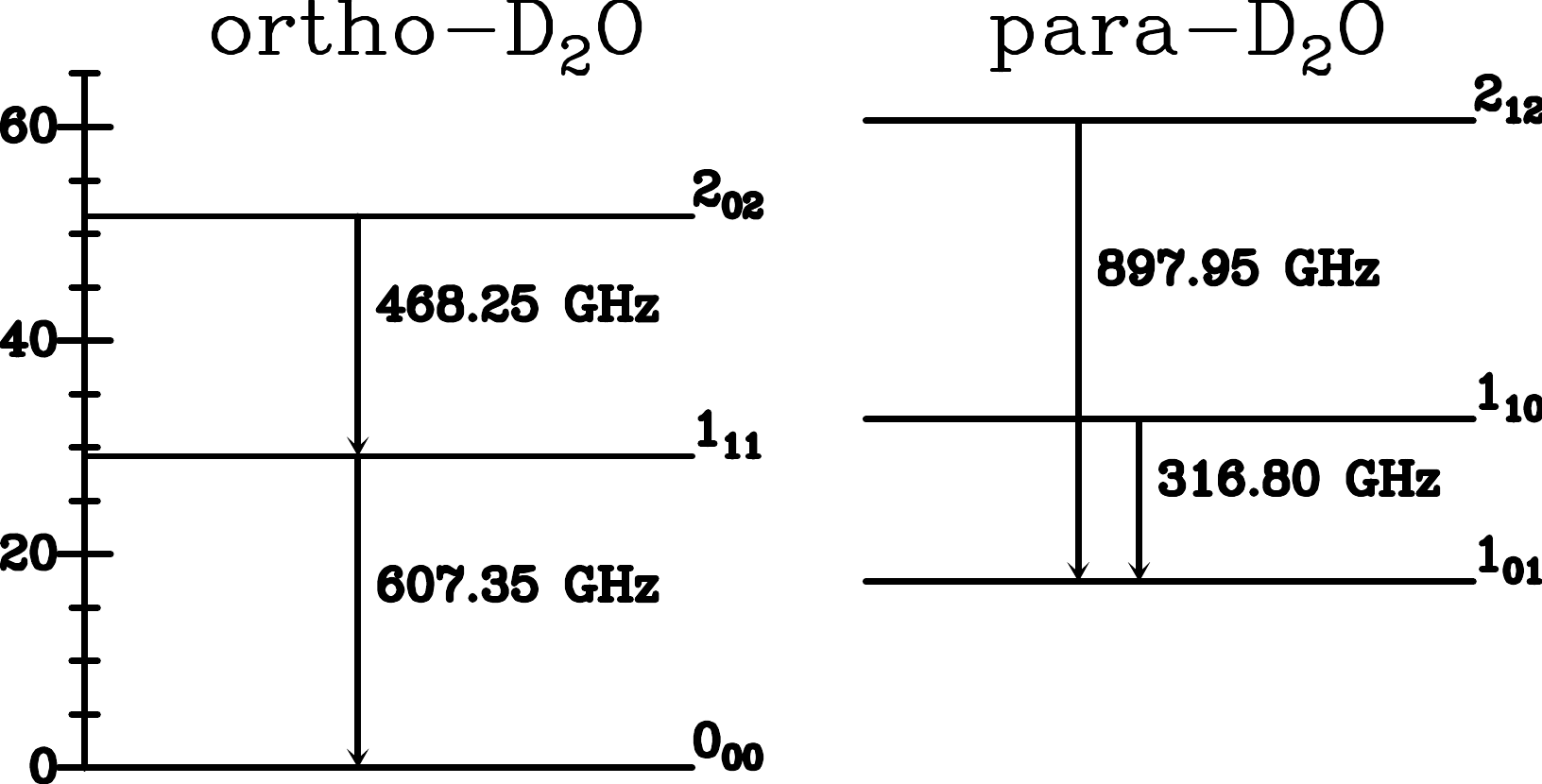}
  \caption{Energy levels for the lowest rotational transitions of ortho-- and para--D$_2$O.}
  \label{transitions}
\end{figure}

\begin{table*}[t!]
  \caption{Derived parameters of the absorbing components of the ortho-- and para--D$_2$O fundamental
    lines. } %Note that the parameters are in T$_a^*$ for ortho--D$_2$O and T$_{mb}$ for para--D$_2$O (see text).}    
\label{table1}      % is used to refer this table in the text
\centering                          % used for centering table
\begin{tabular}{@{}c@{~}c@{~~~}c@{~~~}c c c c c c c c@{}}        
\hline\hline                 % inserts double horizontal lines
Species      &  Transition & Frequency & $A_{\rm ij}$  & $E_{\rm up}$ & Telescope & $\int T_{\rm abs} dv$ & $T_{\rm abs}^{}$ & $FWHM$ &$V_{LSR}$ & $|T_{\rm abs}|^{\rm}$/$T_{\rm C}$  \\  %& $T_C$ & $\tau$ \\    
	           &                     & (GHz)            &     (s$^{-1}$)             &    (K)          &     & (mK km/s)   & (mK) & (km/s)  & (km/s) \\ % & (mK) & \\
\hline                        % inserts single horizontal line
ortho--D$_2$O & 1$_{1,1}$--0$_{0,0}$ &   607.34945  & 2.95 $\times$ 10$^{-3}$& 29.1 & HIFI 1b/HRS & -90 $\pm$ 8 & -199 $\pm$ 8  & 0.43 $\pm$ 0.02 & 4.22 $\pm$ 0.01 & 0.64 $\pm$ 0.03 \\%&  234 $\pm$ 19 & X $\pm$ X\\
para--D$_2$O  & 1$_{1,0}$--1$_{0,1}$ &   316.79981  & 6.29 $\times$ 10$^{-4}$ & 32.6 & JCMT  & -120 $\pm$ 49 & -220 $\pm$ 30 & 0.55 $\pm$ 0.15 & 4.15 $\pm$ 0.04 & 0.57 $\pm$ 0.08 \\% & 850 $\pm$ 35 & X $\pm$ X\\
para--D$_2$O  & 2$_{1,2}$--1$_{0,1}$ &   897.94711  & 8.61 $\times$ 10$^{-3}$  & 60.5 & HIFI  3b/WBS & -73 $\pm$ 13 & -167 $\pm$ 12 & 0.41 $\pm$ 0.04 & 4.04 $\pm$ 0.02 & 0.22 $\pm$ 0.02 \\ %  & X $\pm$ X \\
\hline                                   %inserts single line
\end{tabular}
\end{table*}

The parameters of the lines, taking into account the ortho--
and para--D$_2$O forms (within the CASSIS software\footnote{Developed by 
IRAP-UPS/CNRS: \url{http://cassis.irap.omp.eu}}) separated from the Cologne Database for
Molecular Spectroscopy \citep{muller2005,brunken2007}, are reported in Table \ref{table1}.  
The fundamental ortho--D$_2$O transition at 607.349 GHz is clearly detected in absorption against the strong
continuum ($\sim$ 0.3 K). %, at the velocity of $\sim$4.2 km\,s$^{-1}$ and shows a FWHM about 0.43 km\,s$^{-1}$. 
We present here the first tentative detection of the para--D$_2$O 2$_{1,2}$--1$_{0,1}$
transition at 897.95 GHz, seemingly in absorption against the continuum ($\sim$ 0.75 K). In the same table, we also report the parameters of the 
para--D$_2$O 1$_{1,0}$--1$_{0,1}$ fundamental line previously observed at  316.800 GHz with the JCMT by \citet{butner2007}. We used the line decomposition in three Gaussian components derived by \citet{vastel2010} : two of them are attributed to the broad emission component and the deep absorbing line of D$_2$O, whereas an additional component emitted at 10.1 km\,s$^{-1}$ is likely due to the 7$_0$--6$_0$+ CH$_3$OD line at 316.795 GHz. This latter component was subtracted in Figure \ref{observations}.
The 2$_{0,2}$--1$_{1,1}$ ortho transition at 468.25 GHz (see Fig. \ref{transitions}) has not been observed, and we consequently only provide a prediction with the modeling.

Observations were recently carried out with the Atacama Pathfinder EXperiment (APEX) telescope between 270 and 330 GHz to complete The IRAS16293-2422 Millimeter and Submillimeter Spectral Survey \citep[TIMASSS;][]{caux2011}. While the para--D$_2$O fundamental 1$_{1,0}$--1$_{0,1}$ line at 317 GHz is also detected in this survey, it shows a much lower signal-to-noise ratio than the one observed with JCMT \citep{butner2007}. This observation is nevertheless useful due to the accurate continuum at this frequency (0.235 K in $T_A^*$). %, see Sect. \ref{modeling}). 
The beam efficiency is 0.73 and the forward efficiency 0.97.

\section{Modeling}
\label{modeling}

In order to determine the D$_2$O/HDO and D$_2$O/H$_2$O ratios within the different components of this source 
(hot corino, colder envelope, and external absorption layer), we developed the same modeling as performed in 
\citet{coutens2012} for the HDO and H$_2$$^{18}$O species with the RATRAN radiative transfer code \citep{ratran}. We employed the ortho-- and para--D$_2$O collisional coefficients calculated with ortho-- and para--H$_2$ by \citet{faure2012} in the 5--100\,K 
range. We assumed that the ortho-to-para D$_2$O ratio is 2:1, 
compatible with the results from \citet{vastel2010}. Using the same parameters for the modeling (see details in \citealp{coutens2012}), 
we realized that 
the continuum modeled at 317 GHz (0.38 K) is not consistent with the continuum observed by \citet{butner2007} 
at 0.85 K. To determine whether this divergence comes from the modeling or from the 
observations, we used the recent APEX observations. 
The SSB continuum observed with APEX is 0.31\,K ($T_{\rm mb}$). Our modeling predicts
a 0.29 K continuum value at the same frequency. From the comparison between transitions from overlapping 
frequency ranges, we can estimate the calibration to be 20\%. The value estimated from the modeling is therefore 
consistent with the APEX continuum value, and in the following we will use the 0.38 K predicted by our modeling for the continuum 
of the 317 GHz transition observed with JCMT. 

\begin{figure*}[!t]
  \centering
    \includegraphics[width=17.0cm]{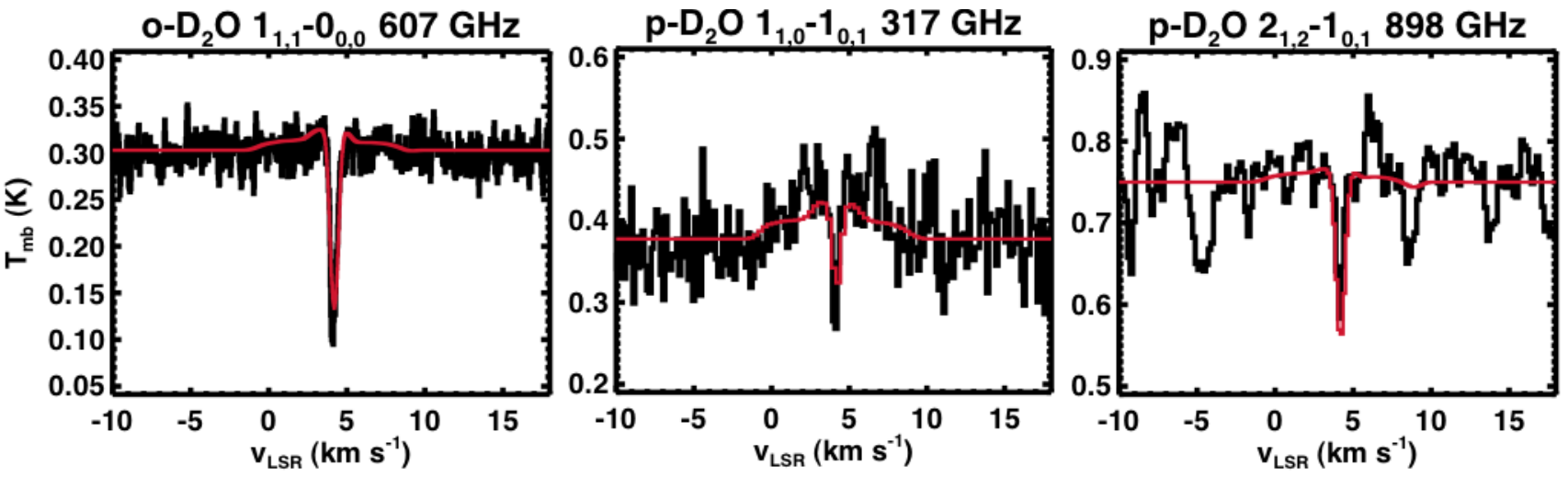}
  \caption{In \textit{black}: D$_2$O transitions observed with HIFI and JCMT. The CH$_3$OD (7$_0$--6$_0$+) transition at 316.795074 GHz contaminating the para--D$_2$O (1$_{1,0}$--1$_{0,1}$) line profile 
  has been subtracted for a direct comparison with the D$_2$O best-fit model. In \textit{red}: best-fit model obtained when adding an absorbing layer with
  N(para--D$_2$O) = 1.1 $\times$ 10$^{12}$ cm$^{-2}$ and  N(ortho--D$_2$O) = 1.4 $\times$ 10$^{12}$ cm$^{-2}$. The resulting inner abundance is 
  7 $\times$ 10$^{-10}$ and the outer abundance is 5 $\times$ 10$^{-12}$. }
  \label{observations}
\end{figure*}

With the source structure (hot corino + colder envelope) derived by \citet{crimier2010}, the modeling is not able to reproduce any absorption 
feature for the D$_2$O fundamental transitions. The overall absorption can only be reproduced  
by the external absorbing layer, which was discovered by \citet{coutens2012} in their efforts to reproduce the absorption components seen on the HDO fundamental transitions. Assuming an H$_2$ density of $\sim$ 10$^3$--10$^4$ cm$^{-3}$, a kinetic temperature of $\sim$ 10--30 K and 
a b-Doppler parameter of 0.3 km\,s$^{-1}$ (turbulence), the ortho--D$_2$O (respectively para--D$_2$O) 
column density in this layer is about 1.4 $\times$ 10$^{12}$ cm$^{-2}$ (respectively 1.1 $\times$ 10$^{12}$ cm$^{-2}$). Our 
modeling predicts that the 2$_{1,2}$--1$_{0,1}$ para--D$_2$O transition is expected in absorption as well, tracing the 
external absorbing layer. The tentatively detected absorbing component at 898 GHz is consequently in perfect agreement with the modeling (see Fig. \ref{observations}).
The kinetic temperature ($\sim$ 10--30 K) has a limited influence on the absorbing transitions, but hydrogen density 
seems to have more influence. With a 10$^5$ cm$^{-3}$ density, the absorption depth is less pronounced for the 
fundamental 1$_{1,0}$--1$_{0,1}$ and 1$_{1,1}$--0$_{0,0}$ transitions at 317 and 607 GHz, requiring a higher D$_2$O column density. However, the higher energy para--D$_2$O  transition 
absorption is overestimated compared to the observations, suggesting a lower H$_2$ density. 
From the above estimations, the D$_2$O ortho-to-para ratio is, in the absorbing layer, about 1.3, close to the value determined by \citet{vastel2010}. 
However, taking into account the observation and modeling uncertainties, we cannot exclude that the measured value might be consistent with the 2:1 value at equilibrium. 
The quoted column densities are slightly higher (less than a factor 2) than those estimated by \citet{vastel2010}, 
but within their 3$\sigma$ uncertainties. 
If this absorbing layer is produced by photodesorption at A$_V$ $\sim$ 1--4 mag as predicted by \citet{Hollenbach2009}, the D$_2$O abundance in this layer should be about 6.6 $\times$ 10$^{-10}$--2.7 $\times$ 10$^{-9}$, using the relation between the H$_2$ column density and the visual extinction, N(H$_2$)/A$_V$ = 9.4 $\times$ 10$^{20}$ cm$^{-2}$ mag$^{-1}$ \citep{Frerking1982}.

\begin{figure}[!t]
  \centering
  \includegraphics[width=9cm]{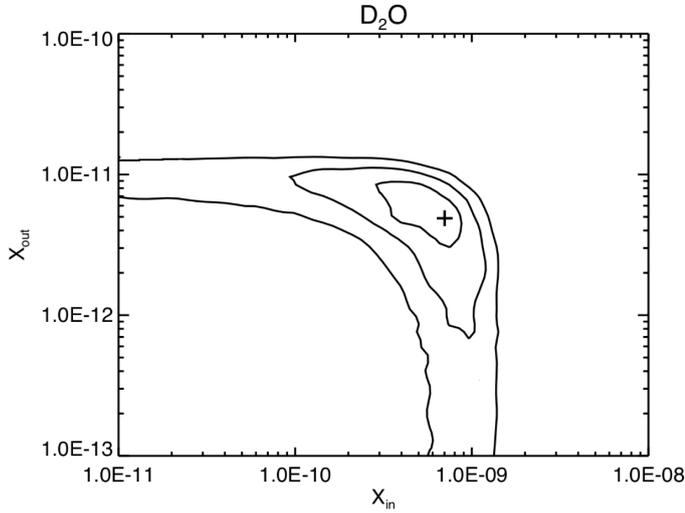}
  \caption{$\chi^2$ contours at 1$\sigma$, 2$\sigma$, and 3$\sigma$ obtained when adding an absorption layer with a para--D$_2$O column 
  density of 1.1 $\times$ 10$^{12}$ cm$^{-2}$ and ortho--D$_2$O column density of 1.4 $\times$ 10$^{12}$ cm$^{-2}$. The best-fit model is 
  represented by the symbol ``+".}
  \label{chi2}
\end{figure}

To estimate the D$_2$O abundances in the hot corino ($X_{\rm in}$) and the outer envelope ($X_{\rm out}$) of IRAS~16293, we ran a grid of models that take into account the structure of \citet{crimier2010} as well as the foregound absorbing layer previously defined.
The best-fit modeling is then obtained for an internal abundance $X_{\rm in}$ = 7 $\times$ 10$^{-10}$ 
and an external abundance $X_{\rm out}$ = 5 $\times$ 10$^{-12}$ from the $\chi^2$ minimization (see Fig. \ref{chi2})
computed on the line profiles according to the following formalism:
\begin{equation}
\label{eq_chi2}
 \chi^2 = \sum_{i=1}^{N} \sum_{j=1}^{ n_{\rm chan}} \frac{(T_{\textrm{obs},ij}-T_{\textrm{mod},ij})^2}{({rms}_{i})^2},
\end{equation}
with $N$ the number of lines $i$, $n_{\rm chan}$ the number of channels $j$ for each line, $T_{\rm obs,ij}$ and $T_{\rm mod,ij}$ the intensity observed and predicted by the RATRAN non-LTE modeling respectively in the channel $j$ of the line $i$ and $rms_{ \rm i}$ the rms of the line $i$. 
Here, the reduced $\chi^2$ is equal to 2.5.
At 2$\sigma$, the inner and outer abundances ranges are $X_{\rm in}$ = [9 $\times$ 10$^{-11}$ -- 1.1 $\times$ 10$^{-9}$] and $X_{\rm out}$ = [8 $\times$ 10$^{-13}$ -- 1.1 $\times$ 10$^{-11}$].
Only upper limits can, however, be given at 3$\sigma$, with $X_{\rm in}$ $\le$ 1.3 $\times$ 10$^{-9}$ and 
$X_{\rm out}$ $\le$ 1.3 $\times$ 10$^{-11}$. 
The best-fit modeling is shown in Figure \ref{observations}.

To check the influence on the ortho-to-para ratio assumed for D$_2$O in the protostellar envelope (i.e., the equilibrium value of 2:1), we ran another grid of models with an ortho-to-para D$_2$O ratio equal to 1.3 in the envelope, i.e., the same as estimated in the absorbing layer. The results in both cases are quite in agreement and do not allow us to conclude with an ortho-to-para ratio different from the equilibrium value 2:1. In addition, we checked that the para--D$_2$O emission line predicted at 317 GHz was consistent with the line observed with APEX.
Regarding the 2$_{0,2}$--1$_{1,1}$ ortho transition at 468 GHz, we predict that the line only shows an emission component with an intensity about 0.1 K ($T_{\rm mb}$) if observed with APEX.

\begin{table*}[ht]
%\vspace{-1cm}
\caption{HDO/H$_2$O, D$_2$O/HDO, and D$_2$O/H$_2$O abundance ratios estimated towards the protostar IRAS~16293.}
%\vspace{0.5cm}
\label{resume2} 
%\centering
%\begin{footnotesize}
\begin{center}                        % used for centering table
\begin{tabular}{c |c |c| c| c |c }
\hline
\hline	
& \multicolumn{2}{c|}{Hot corino} & \multicolumn{2}{c|}{Outer envelope} & \multicolumn{1}{c}{Photodesorption layer} \\
\cline{2-6}
& Best-fit & 3$\sigma$ &  Best-fit & 3$\sigma$ &  A$_V$ $\sim$ 1 -- 4 mag\\
\hline
HDO$^{a}$ & 1.8 $\times$ 10$^{-7}$ & 1.4 -- 2.4 $\times$ 10$^{-7}$ &  8  $\times$ 10$^{-11}$ & 5.5 -- 10.6 $\times$ 10$^{-11}$ & $\sim$ 0.6 -- 2.4  $\times$ 10$^{-8}$ \\
H$_2$O$^{a,b}$ & 1 $\times$ 10$^{-5}$ & 4.7 -- 40.0 $\times$ 10$^{-6 }$& 1.5 $\times$ 10$^{-8}$ & 7.0 -- 22.5 $\times$ 10$^{-9}$ & $\sim$ 1.3 -- 5.3 $\times$ 10$^{-7}$ \\
D$_2$O & 7 $\times$ 10$^{-10}$ & $\le$ 1.3 $\times$ 10$^{-9}$ & 5 $\times$ 10$^{-12}$ & $\le$ 1.3 $\times$ 10$^{-11}$ & $\sim$ 6.6 -- 27 $\times$ 10$^{-10}$ \\
HDO/H$_2$O & 1.8\% & 0.4\% -- 5.1\% & 0.5\% & 0.3\% -- 1.5\% & $\sim$ 4.8\%$^c$\\
D$_2$O/HDO & 0.4\% & $\le$ 0.9\% & 6.3\% & $\le$ 23\% & $\sim$ 10.8\%$^c$\\
D$_2$O/H$_2$O & 0.007\% & $\le$ 0.03\% & 0.03\% & $\le$ 0.2\% & $\sim$ 0.5\%$^c$\\
\hline 
\end{tabular}
\end{center} 
%\end{footnotesize}
%\vspace{-0.3cm}
$^{a}$ \small{The HDO and H$_2$O abundances from \citet{coutens2012} were recalculated with the $\chi^2$ minimization shown in Equation \ref{eq_chi2}.} 
$^{b}$ \small{The H$_2$O abundances are obtained from the H$_2$$^{18}$O data using the collision rates with H$_2$ calculated by \citet{daniel2011}. The H$_2$$^{16}$O/H$_2$$^{18}$O} ratio is assumed to be 500.
$^{c}$ \small{The ratios remain valid if the absorbing layer is not due to the photodesorption.}
\end{table*}

\section{Discussion}

\subsection{Ortho-to-para D$_2$O ratio}
At the thermal equilibrium, the ortho-to-para ratio of D$_2$O is higher or equal to 2, depending on temperature (see Fig. 4 in \citealt{vastel2010}). If the ortho-to-para D$_2$O ratio estimated at $\sim$ 1.3 in the absorbing layer really deviates from the 2:1 value as already suggested by \citet{vastel2010}, this could mean that some nuclear spin exchanges occur in the gas phase or at the grain surface. 
Indeed, the ortho-to-para ratio should be initially set during the molecule formation at a value reflecting the grain temperature ($\geq$ 2).
Some spin exchange mechanisms would be consequently required to decrease the ortho-to-para ratio. These mechanisms are, however, poorly known.
We can imagine some exchange mechanisms at the grain surface with interaction between electronic or nuclear spins. Unfortunately, the theoretical and laboratory measurements \citep{limbach2006,Pardanaud2007} are difficult to carry out. 
In the gas phase, no fast radiative transition is possible between ortho and para species, except via proton exchange with ions. 
For example, to modify the ortho-to-para ratio of water, collisions with H$^+$ and H$_3$$^+$ are required. % \citep{lis2010}.
For D$_2$O, the following reaction could consequently occur via D exchange between D$_2$O and H$_2$D$^+$: 
\vspace{0.2cm} \\
 {o-D$_2$O }+ H$_2$D$^+$ $\longleftrightarrow$ {p-D$_2$O} + H$_2$D$^+$,
 \vspace{0.2cm} \\
However, this reaction could also lead to other products for which the branching ratios are unknown: 
\vspace{0.2cm} \\
D$_2$O + H$_2$D$^+$ $\longleftrightarrow$ HD$_2$O$^+$ + HD 
\vspace{0.2cm} \\
D$_2$O + H$_2$D$^+$ $\longleftrightarrow$ D$_3$O$^+$ + H$_2$ 
\vspace{0.2cm} \\
This reaction has never been studied experimentally or theoretically, but could be exothermic like the H$_2$O + H$_3^+$ reaction \citep{Woon2009}.
The nuclear spin conversion could also be possible thanks to an exchange with D$^+$.

\subsection{Water deuterium fractionation}
As already noticed in \citet{coutens2012}, the HDO/H$_2$O ratios determined in the hot corino and in the external absorbing layer are 
similar ($\sim$2--5\%), although the corresponding densities are very different, a few 10$^8$ cm$^{-3}$ in the 
hot corino region and, typical of a molecular cloud, probably 10$^3$--10$^4$ cm$^{-3}$ in the external layer. Water shows a different behavior from other tracers as methanol (CH$_3$OH) 
and formaldehyde (H$_2$CO), whose deuteration increases with the density. Indeed, the deuteration of these heavier molecules that need CO ices to form  is correlated with the CO depletion (linked to the molecular hydrogen density: \citealp{caselli1999,bacmann2002}) in prestellar cores \citep{bacmann2003,bacmann2007}.
In contrast, the water fractionation ratio does not seem to vary with density. As discussed 
in \citet{coutens2012}, it is therefore possible that water was formed at low densities in the early stages
of the star formation before the protostellar collapse. 

\begin{figure*}[!t]
  \centering
      \includegraphics[width=9cm]{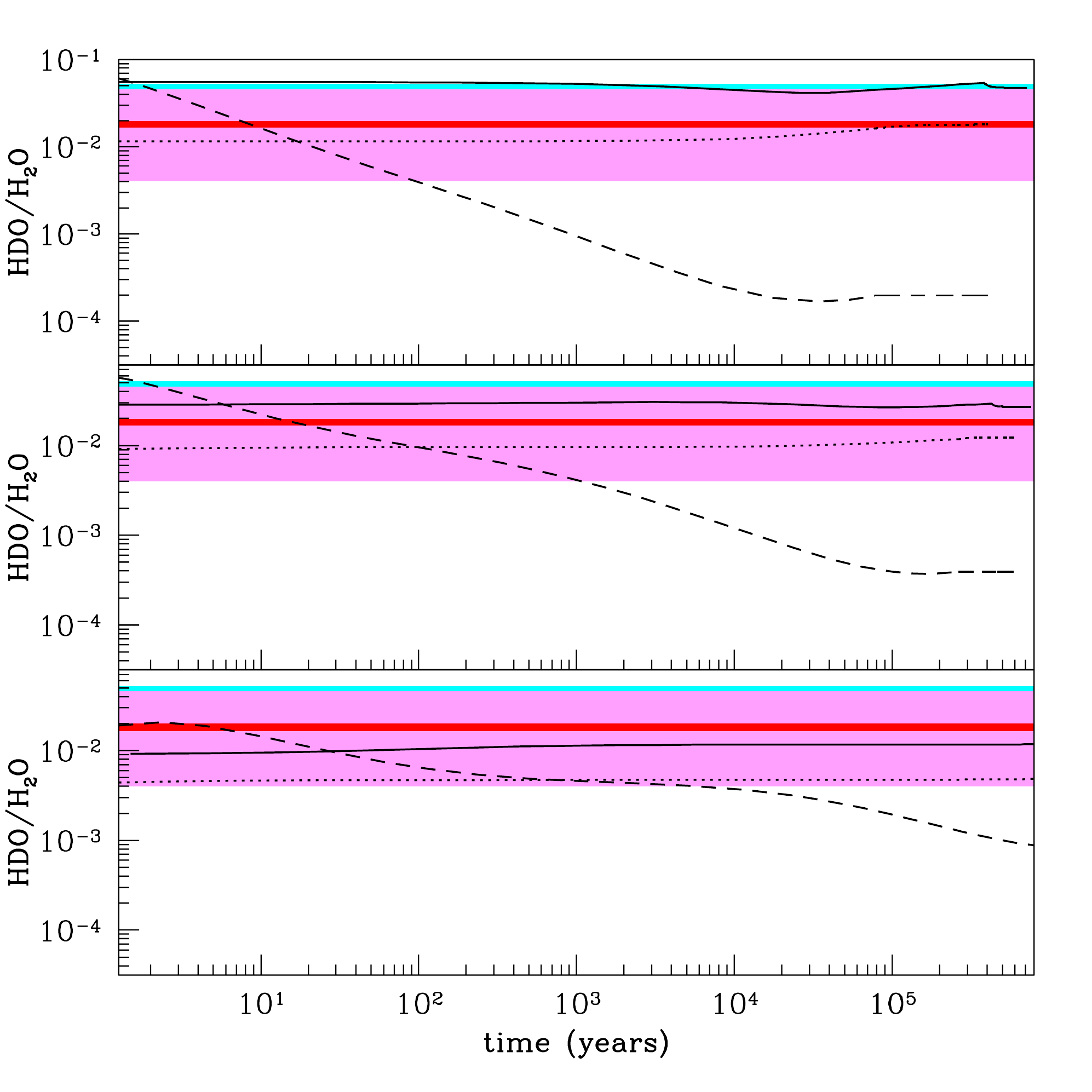}
      \includegraphics[width=9cm]{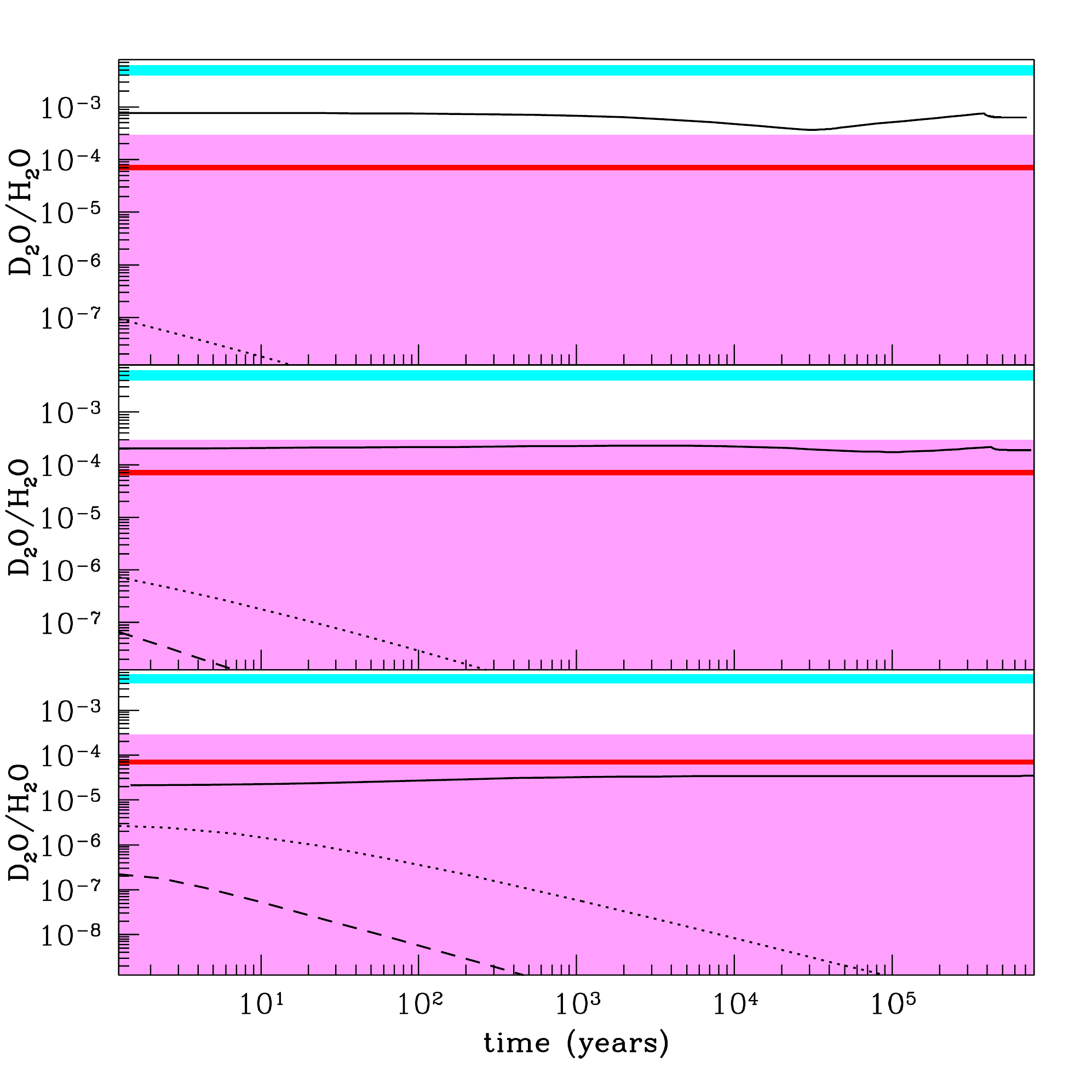}
  \caption{Deuterium fractionation of water predicted by the grain surface chemical model (see text) assuming 3 initial dust temperatures $T_{\rm dust}$ = 12 K (dashed lines), 
  15 K (dotted lines) and 20 K (solid lines) and densities $n$ = 10$^3$ (lower panel), 10$^4$ (middle panel) and 5 $\times$ 10$^4$ cm$^{-3}$ (upper panel).
  The HDO/H$_2$O and D$_2$O/H$_2$O ratios determined at 3$\sigma$ in the hot corino are indicated in pink with a red line for the best-fit value. The ratios of the absorbing layer are shown with a blue line.}
  \label{model}
\end{figure*}

To understand how the deuterium fractionation of water depends on the initial conditions of the cloud in which ices are forming, we used in the present study a grain surface chemical model \citep{cazaux2011}, in which we considered the 
formation of ices by accreting species from the gas phase. The icy mantles then grow until 
99\% of CO and O from the gas are depleted onto dust.
We set the initial densities of O and CO in the gas similar to densities of O and CO in translucent clouds, i.e., $n$(O) $\sim$ $n$(CO) $\sim$ 1.5 $\times$ 10$^{-4}$ $n_{\rm H}$ \citep{Hollenbach2009}. 
The gas phase species accrete on the dust and the chemical evolution of ices can be followed with time until most of O and CO are depleted from the gas phase ($t$ $\sim$ 1--5 $\times$ 10$^5$ years, depending on the density of the medium). During the formation of ices, O and CO gas phase abundances are decreasing while the abundances of gas phase H$_2$, H and D are kept constant.  
The density of the cloud sets the gas phase D/H ratio though ion-molecule chemistry (deuteration of H$_3^+$ sets the gas D/H ratio, as described in \citealp{roberts2002}). Therefore, a translucent cloud with an initial density of 10$^3$, 10$^4$, and 5 $\times$ 10$^4$ cm$^{-3}$ results in a D/H ratio in the gas phase of 5 $\times$ 10$^{-3}$, 5 $\times$ 10$^{-2}$, and 10$^{-2}$ respectively \citep{cazaux2011}. 
Consequently, we assume these three initial 
values for the density (10$^3$,  10$^4$, and 5 $\times$ 10$^4$ cm$^{-3}$) as well as three different temperatures (12, 15, and 20 K). The HDO/H$_2$O as well as the D$_2$O/H$_2$O ratios obtained during the formation of ices are 
presented in Figure \ref{model}. Both ratios are strongly sensitive to the dust temperature. 
Indeed, at low temperatures, H$_2$ is much more abundant than H on the dust grains, and the successive hydrogenations of 
O with H$_2$ dominate the formation of H$_2$O. 
The reaction O($^3$P) + H$_2$ is energetically not accessible at low temperatures since the reaction has a barrier of 0.57 eV and is endo-ergic with an energy of 0.1 eV for H$_2$(v=0) \citep{weck2006}.
However, this reaction becomes exo-ergic for v$>$0 and the barrier decreases to $\sim$0.4 eV for v=1 \citep{Sultanov2005}. The reaction O($^3$P) + H$_2$(v=1) proceeds through tunneling at low temperatures, as shown by \citet{weck2006}. In our model, we consider only the reaction of H$_2$ vibrationally excited with O($^3$P). For a population in equilibrium at a temperature $T_{\rm dust}$, the fraction of molecules in the $J$ = 1 state is $\sim$ $\exp{(-150/T_{\rm dust})}$ $\sim$ 3 $\times$ 10$^{-6}$ -- 5.5 $\times$ 10$^{-4}$ in the range of $T_{\rm dust}$ considered in this study (12 -- 20 K). The reaction through tunneling between H$_2$(v=1) and O($^3$P) dominates at very low temperatures when the dust is saturated with H$_2$ molecules and the oxygen atoms present on the surface repetitively collide with H$_2$.
On the other hand, HDO forms through addition of atomic deuterium. These different formation routes imply that the HDO/H$_2$O ratio scales with D/H$_2$ ratio present in the gas, while the D$_2$O/H$_2$O scales with (D/H$_2$)$^2$.
The reaction between O and H, however, dominates at 12 K at short timescale because of the relatively high initial H/H$_2$ ratio ($\sim$ 10$^{-4}$). However, the H/H$_2$ ratio decreases with time, since H atoms are quickly converted into OH and H$_2$, and the reaction O + H$_2$ dominates (for $t$ $\sim$ 10$^4$ years at $n_{\rm H}$ = 10$^3$ cm$^{-3}$ and for $t$ $\sim$ 10$^5$ years at $n_{\rm H}$ = 10$^4$ cm$^{-3}$).
If the grain is warm, the H$_2$ molecules evaporate and the formation of H$_2$O is insured by successive hydrogenation with H atoms.  Since the formation of HDO involves atomic deuterium, the HDO/H$_2$O ratio becomes comparable to the D/H ratio and the D$_2$O/H$_2$O ratio comparable to (D/H)$^2$. Therefore, the variation of the deuterium fractionation with the initial dust temperature is due to the different routes to form water molecules. 
In their modeling, \citet{cazaux2011} found that the deuterium fractionation of the water ice mantle is not altered by the cloud collapse, pointing to the fact that the deuterium fractionation of water must have been set during the formation of ices.

The HDO/H$_2$O and D$_2$O/H$_2$O observed ratios suggest that the ices present in IRAS 16293 and desorbing in the hot corino were made in a warm ($\sim$ 15 -- 20 K) and dense ($\sim$ 10$^4$ -- 5 $\times$ 10$^4$ cm$^{-3}$) translucent cloud.
Recently, \citet{Taquet2012} modeled the formation of deuterated water ice in cold conditions with a pseudo time-dependent multilayer approach (see \citealt{Taquet2012} for a full description of the model). They showed that the relatively high deuterium fractionation of water could be attributed to the formation of water ices in dark phases (either at $n_{\rm H}$ $\sim$ 10$^4$ cm$^{-3}$ at low temperatures, $\sim$ 10 K, and high visual extinction, larger than 4 magnitudes, or for higher densities, $\sim$ 10$^5$ cm$^{-3}$, at higher temperatures), where the D/H ratio starts to efficiently increase.
Although the chemical models diverge on the deuteration processes which are enhanced for warm dust in one case \citep{cazaux2011} and for cold dust in the other case \citep{Taquet2012}, our results agree that the high deuterium fractionation originates from a dense medium (both models need dense medium for high D/H ratio).

In the extended absorbing layer, which is possibly due to photodesorption mechanisms, the observed D$_2$O/H$_2$O ratio appears much larger compared to the hot corino value, in contrast to the HDO/H$_2$O ratio.
It is also higher than all the chemical predictions.
If the D$_2$O/H$_2$O ratio observed in the gas phase of this layer really reflects the D$_2$O/H$_2$O ratio at the grain surface, 
it would consequently suggest that water formed in a denser ($\gtrsim$ 5 $\times$ 10$^4$ cm$^{-3}$) region with warm ($\sim$ 20 K) gas. A high density (10$^5$--10$^6$ cm$^{-3}$) is also required for the chemical modeling by \citet{Taquet2012}.
However, these results disagree with our observational constraints obtained from the density of this absorbing layer ($<$ 10$^5$ cm$^{-3}$, see Sect. \ref{modeling}).
Nevertheless, we cannot exclude that, in this external layer, water can also be formed by gas phase mechanisms through ion-molecule reactions. 
This could also be one of the possible explanations of the difference between the D$_2$O/H$_2$O ratios determined in the hot corino and in the absorbing layer. Other reasons can be mentioned to explain the discrepancy of the water deuterium fractionation between the hot corino and the absorbing layer.
During the heating of the grain surfaces, an isotope exchange between water molecules may also occur at 150 K \citep{Smith1997, Dulieu2010}, leading to a decrease of the D$_2$O abundance in the hot corino region.
Also, in the hot corino, large columns of H$_2$O, HDO, and D$_2$O could self-shield these molecules. Since H$_2$O is more abundant than its deuterated forms, it should self-shield first, reducing its photodissociation. Then if important columns of HDO and D$_2$O can be reached, these molecules will be shielded. This implies a stratification of molecules and should favor the amount of water compared to its deuterated forms in environments where lots of water is present in the gas phase.

Recently, \citet{persson2012} used interferometric (SMA and ALMA) observations of one HDO transition (226 GHz) and two H$_2$$^{18}$O transitions (203 and 692 GHz) to determine the water deuterium fractionation in the hot corino of  IRAS~16293. Water is only emitted by the core A and not by its binary component B. With their different method,  \citet{persson2012} derive a fractionation ratio about a factor of 10 lower than ours \citep{coutens2012}. The origin of the divergence between these results is not clear, but could arise from the uncertain inner structure of the source.
Indeed, a recent publication by \citet{pineda2012} mentions that ALMA observations towards the source A would be consistent with the rotation of a disk, which is close to being edge-on. None of the two studies assume such a complex structure in their modeling, maybe implying the divergence between the estimated HDO/H$_2$O ratios.
Compared with the predictions of the chemical model by \citet{cazaux2011}, the ratio determined by \citet{persson2012} would require a low initial dust temperature (around 12 K), in contrast to our results.
But doubly deuterated water has never been detected with interferometers, making it not possible to check whether the D$_2$O/H$_2$O ratio is also in agreement with an origin on cold dust grains and provide more constraints on the initial density.
Consequently, complementary D$_2$O observations at high spatial resolution would be necessary to assess precisely the conditions in which ices are forming with a method based on interferometric data as in \citet{persson2012}.

\section{Conclusions}

Thanks to the numerous HDO and H$_2$$^{18}$O transitions observed with 
the HIFI instrument on board the \textit{Herschel} observatory, we were able \citep{coutens2012} to 
constrain the abundances of these species from the hot corino throughout the envelope of the IRAS 16293 
protostar to the external absorbing layer. In the present study, we have used the 
same non-LTE radiative transfer modeling to constrain the D$_2$O abundances, employing the 
previous detections of the para and ortho fundamental transitions \citep{butner2007,vastel2010} as well as the tentative detection 
of the higher energy para--D$_2$O 2$_{1,2}$--1$_{0,1}$ transition at 898 GHz in Herschel/HIFI band 3b. 
To estimate the abundances, we used the full set of D$_2$O collisional rate coefficients computed with ortho-- and para--H$_2$ 
\citep{faure2012}, which were not available for the original detections.
The results are presented in Table 2. The best-fit inner abundance $X_{\rm in}$(D$_2$O) is about 7 $\times$ 10$^{-10}$, 
whereas the best-fit outer abundance $X_{\rm out}$(D$_2$O) is about 5 $\times$ 10$^{-12}$. 
The absorbing D$_2$O components cannot be reproduced without taking into account the external absorbing layer.
The column density in this layer is about 2.5 $\times$ 10$^{12}$ cm$^{-2}$, leading to an abundance ranging from 6.6 $\times$ 10$^{-10}$ to 2.7 $\times$ 10$^{-9}$ if the lines are produced in
a photodesorption layer at $A_V$ $\sim$ 1--4 magnitudes. % ) ranges from 6.6 to 27 $\times$ 10$^{-10}$. 
We then compared the HDO/H$_2$O and D$_2$O/H$_2$O observed ratios with the predictions of a grain surface chemical model 
\citep{cazaux2011}, in which we considered the formation of ices by accreting species from the gas phase.
Our estimations of the HDO/H$_2$O and D$_2$O/H$_2$O ratios suggest that
the ices present in IRAS 16293 and desorbing in the hot corino were made in a warm ($\sim$ 15 -- 20 K) and dense 
(10$^{4}$ -- 5 $\times$ 10$^{4}$ cm$^{-3}$) translucent cloud. The observed D$_2$O/H$_2$O ratio appears much larger in the absorbing 
layer than in the hot corino, and we explored tentative explanations.

\begin{acknowledgements}
  HIFI was designed and built by a consortium of institutes and
  university departments from across Europe, Canada, and the United
  States under the leadership of SRON Netherlands Institute for Space
  Research, Groningen, The Netherlands and with major contributions
  from Germany, France and the US. Consortium members are: Canada:
  CSA, U.Waterloo; France: IRAP (formerly CESR), LAB, LERMA, IRAM; Germany: KOSMA,
  MPIfR, MPS; Ireland, NUI Maynooth; Italy: ASI, IFSI-INAF,
  Osservatorio Astrofisico di Arcetri-INAF; Netherlands: SRON, TUD;
  Poland: CAMK, CBK; Spain: Observatorio Astron\'omico Nacional (IGN),
  Centro de Astrobiolog\'{\i}a (CSIC-INTA). Sweden: Chalmers
  University of Technology - MC2, RSS \& GARD; Onsala Space
  Observatory; Swedish National Space Board, Stockholm University -
  Stockholm Observatory; Switzerland: ETH Zurich, FHNW; USA: Caltech,
  JPL, NHSC.  We thank the CNES (Centre National d'Etudes Spatiales) for its financial support.
  VW thanks the French National Program PCMI for the partial funding of her research.
\end{acknowledgements}

\bibliographystyle{aa}
\bibliography{biblio}

\begin{thebibliography}{41}
\expandafter\ifx\csname natexlab\endcsname\relax\def\natexlab#1{#1}\fi

\bibitem[{{Bacmann} {et~al.}(2010){Bacmann}, {Caux}, {Hily-Blant}, {Parise},
  {Pagani}, {Bottinelli}, {Maret}, {Vastel}, {Ceccarelli}, {Cernicharo},
  {Henning}, {Castets}, {Coutens}, {Bergin}, {Blake}, {Crimier}, {Demyk},
  {Dominik}, {Gerin}, {Hennebelle}, {Kahane}, {Klotz}, {Melnick}, {Schilke},
  {Wakelam}, {Walters}, {Baudry}, {Bell}, {Benedettini}, {Boogert}, {Cabrit},
  {Caselli}, {Codella}, {Comito}, {Encrenaz}, {Falgarone}, {Fuente},
  {Goldsmith}, {Helmich}, {Herbst}, {Jacq}, {Kama}, {Langer}, {Lefloch}, {Lis},
  {Lord}, {Lorenzani}, {Neufeld}, {Nisini}, {Pacheco}, {Pearson}, {Phillips},
  {Salez}, {Saraceno}, {Schuster}, {Tielens}, {van der Tak}, {van der Wiel},
  {Viti}, {Wyrowski}, {Yorke}, {Faure}, {Benz}, {Coeur-Joly}, {Cros},
  {G{\"u}sten}, \& {Ravera}}]{bacmann2010}
{Bacmann}, A., {Caux}, E., {Hily-Blant}, P., {et~al.} 2010, \aap, 521, L42

\bibitem[{{Bacmann} {et~al.}(2002){Bacmann}, {Lefloch}, {Ceccarelli},
  {Castets}, {Steinacker}, \& {Loinard}}]{bacmann2002}
{Bacmann}, A., {Lefloch}, B., {Ceccarelli}, C., {et~al.} 2002, \aap, 389, L6

\bibitem[{{Bacmann} {et~al.}(2003){Bacmann}, {Lefloch}, {Ceccarelli},
  {Steinacker}, {Castets}, \& {Loinard}}]{bacmann2003}
{Bacmann}, A., {Lefloch}, B., {Ceccarelli}, C., {et~al.} 2003, \apjl, 585, L55

\bibitem[{{Bacmann} {et~al.}(2007){Bacmann}, {Lefloch}, {Parise}, {Ceccarelli},
  \& {Steinacker}}]{bacmann2007}
{Bacmann}, A., {Lefloch}, B., {Parise}, B., {Ceccarelli}, C., \& {Steinacker},
  J. 2007, in Molecules in Space and Laboratory

\bibitem[{{Bottinelli} {et~al.}(2004){Bottinelli}, {Ceccarelli}, {Neri},
  {Williams}, {Caux}, {Cazaux}, {Lefloch}, {Maret}, \&
  {Tielens}}]{bottinelli2004}
{Bottinelli}, S., {Ceccarelli}, C., {Neri}, R., {et~al.} 2004, \apjl, 617, L69

\bibitem[{{Br{\"u}nken} {et~al.}(2007){Br{\"u}nken}, {M{\"u}ller}, {Endres},
  {Lewen}, {Giesen}, {Drouin}, {Pearson}, \& {M{\"a}der}}]{brunken2007}
{Br{\"u}nken}, S., {M{\"u}ller}, H.~S.~P., {Endres}, C., {et~al.} 2007,
  Physical Chemistry Chemical Physics, 9, 2103

\bibitem[{{Butner} {et~al.}(2007){Butner}, {Charnley}, {Ceccarelli}, {Rodgers},
  {Pardo}, {Parise}, {Cernicharo}, \& {Davis}}]{butner2007}
{Butner}, H.~M., {Charnley}, S.~B., {Ceccarelli}, C., {et~al.} 2007, \apjl,
  659, L137

\bibitem[{{Caselli} {et~al.}(1999){Caselli}, {Walmsley}, {Tafalla}, {Dore}, \&
  {Myers}}]{caselli1999}
{Caselli}, P., {Walmsley}, C.~M., {Tafalla}, M., {Dore}, L., \& {Myers}, P.~C.
  1999, \apjl, 523, L165

\bibitem[{{Caux} {et~al.}(2011){Caux}, {Kahane}, {Castets}, {Coutens},
  {Ceccarelli}, {Bacmann}, {Bisschop}, {Bottinelli}, {Comito}, {Helmich},
  {Lefloch}, {Parise}, {Schilke}, {Tielens}, {van Dishoeck}, {Vastel},
  {Wakelam}, \& {Walters}}]{caux2011}
{Caux}, E., {Kahane}, C., {Castets}, A., {et~al.} 2011, \aap, 532, A23

\bibitem[{{Cazaux} {et~al.}(2011){Cazaux}, {Caselli}, \& {Spaans}}]{cazaux2011}
{Cazaux}, S., {Caselli}, P., \& {Spaans}, M. 2011, \apjl, 741, L34

\bibitem[{{Ceccarelli} {et~al.}(2010){Ceccarelli}, {Bacmann}, {Boogert},
  {Caux}, {Dominik}, {Lefloch}, {Lis}, {Schilke}, {van der Tak}, {Caselli},
  {Cernicharo}, {Codella}, {Comito}, {Fuente}, {Baudry}, {Bell}, {Benedettini},
  {Bergin}, {Blake}, {Bottinelli}, {Cabrit}, {Castets}, {Coutens}, {Crimier},
  {Demyk}, {Encrenaz}, {Falgarone}, {Gerin}, {Goldsmith}, {Helmich},
  {Hennebelle}, {Henning}, {Herbst}, {Hily-Blant}, {Jacq}, {Kahane}, {Kama},
  {Klotz}, {Langer}, {Lord}, {Lorenzani}, {Maret}, {Melnick}, {Neufeld},
  {Nisini}, {Pacheco}, {Pagani}, {Parise}, {Pearson}, {Phillips}, {Salez},
  {Saraceno}, {Schuster}, {Tielens}, {van der Wiel}, {Vastel}, {Viti},
  {Wakelam}, {Walters}, {Wyrowski}, {Yorke}, {Liseau}, {Olberg}, {Szczerba},
  {Benz}, \& {Melchior}}]{ceccarelli2010}
{Ceccarelli}, C., {Bacmann}, A., {Boogert}, A., {et~al.} 2010, \aap, 521, L22

\bibitem[{{Ceccarelli} {et~al.}(2000){Ceccarelli}, {Castets}, {Caux},
  {Hollenbach}, {Loinard}, {Molinari}, \& {Tielens}}]{ceccarelli2000}
{Ceccarelli}, C., {Castets}, A., {Caux}, E., {et~al.} 2000, \aap, 355, 1129

\bibitem[{{Coutens} {et~al.}(2012){Coutens}, {Vastel}, {Caux}, {Ceccarelli},
  {Bottinelli}, {Wiesenfeld}, {Faure}, {Scribano}, \& {Kahane}}]{coutens2012}
{Coutens}, A., {Vastel}, C., {Caux}, E., {et~al.} 2012, \aap, 539, A132

\bibitem[{{Crimier} {et~al.}(2010){Crimier}, {Ceccarelli}, {Maret},
  {Bottinelli}, {Caux}, {Kahane}, {Lis}, \& {Olofsson}}]{crimier2010}
{Crimier}, N., {Ceccarelli}, C., {Maret}, S., {et~al.} 2010, \aap, 519, A65+

\bibitem[{{Daniel} {et~al.}(2011){Daniel}, {Dubernet}, \&
  {Grosjean}}]{daniel2011}
{Daniel}, F., {Dubernet}, M.-L., \& {Grosjean}, A. 2011, \aap, 536, A76

\bibitem[{{de Graauw} {et~al.}(2010){de Graauw}, {Helmich}, {Phillips},
  {Stutzki}, {Caux}, {Whyborn}, {Dieleman}, {Roelfsema}, {Aarts}, {Assendorp},
  {Bachiller}, {Baechtold}, {Barcia}, {Beintema}, {Belitsky}, {Benz}, {Bieber},
  {Boogert}, {Borys}, {Bumble}, {Ca{\"i}s}, {Caris}, {Cerulli-Irelli},
  {Chattopadhyay}, {Cherednichenko}, {Ciechanowicz}, {Coeur-Joly}, {Comito},
  {Cros}, {de Jonge}, {de Lange}, {Delforges}, {Delorme}, {den Boggende},
  {Desbat}, {Diez-Gonz{\'a}lez}, {di Giorgio}, {Dubbeldam}, {Edwards},
  {Eggens}, {Erickson}, {Evers}, {Fich}, {Finn}, {Franke}, {Gaier}, {Gal},
  {Gao}, {Gallego}, {Gauffre}, {Gill}, {Glenz}, {Golstein}, {Goulooze},
  {Gunsing}, {G{\"u}sten}, {Hartogh}, {Hatch}, {Higgins}, {Honingh}, {Huisman},
  {Jackson}, {Jacobs}, {Jacobs}, {Jarchow}, {Javadi}, {Jellema}, {Justen},
  {Karpov}, {Kasemann}, {Kawamura}, {Keizer}, {Kester}, {Klapwijk}, {Klein},
  {Kollberg}, {Kooi}, {Kooiman}, {Kopf}, {Krause}, {Krieg}, {Kramer},
  {Kruizenga}, {Kuhn}, {Laauwen}, {Lai}, {Larsson}, {Leduc}, {Leinz}, {Lin},
  {Liseau}, {Liu}, {Loose}, {L{\'o}pez-Fernandez}, {Lord}, {Luinge}, {Marston},
  {Mart{\'{\i}}n-Pintado}, {Maestrini}, {Maiwald}, {McCoey}, {Mehdi}, {Megej},
  {Melchior}, {Meinsma}, {Merkel}, {Michalska}, {Monstein}, {Moratschke},
  {Morris}, {Muller}, {Murphy}, {Naber}, {Natale}, {Nowosielski}, {Nuzzolo},
  {Olberg}, {Olbrich}, {Orfei}, {Orleanski}, {Ossenkopf}, {Peacock}, {Pearson},
  {Peron}, {Phillip-May}, {Piazzo}, {Planesas}, {Rataj}, {Ravera}, {Risacher},
  {Salez}, {Samoska}, {Saraceno}, {Schieder}, {Schlecht}, {Schl{\"o}der},
  {Schm{\"u}lling}, {Schultz}, {Schuster}, {Siebertz}, {Smit}, {Szczerba},
  {Shipman}, {Steinmetz}, {Stern}, {Stokroos}, {Teipen}, {Teyssier}, {Tils},
  {Trappe}, {van Baaren}, {van Leeuwen}, {van de Stadt}, {Visser}, {Wildeman},
  {Wafelbakker}, {Ward}, {Wesselius}, {Wild}, {Wulff}, {Wunsch}, {Tielens},
  {Zaal}, {Zirath}, {Zmuidzinas}, \& {Zwart}}]{degraauw2010}
{de Graauw}, T., {Helmich}, F.~P., {Phillips}, T.~G., {et~al.} 2010, \aap, 518,
  L6

\bibitem[{{Dulieu} {et~al.}(2010){Dulieu}, {Amiaud}, {Congiu}, {Fillion},
  {Matar}, {Momeni}, {Pirronello}, \& {Lemaire}}]{Dulieu2010}
{Dulieu}, F., {Amiaud}, L., {Congiu}, E., {et~al.} 2010, \aap, 512, A30

\bibitem[{{Faure} {et~al.}(2012){Faure}, {Wiesenfeld}, {Scribano}, \&
  {Ceccarelli}}]{faure2012}
{Faure}, A., {Wiesenfeld}, L., {Scribano}, Y., \& {Ceccarelli}, C. 2012,
  \mnras, 420, 699

\bibitem[{{Fraser} {et~al.}(2001){Fraser}, {Collings}, {McCoustra}, \&
  {Williams}}]{Fraser2001}
{Fraser}, H.~J., {Collings}, M.~P., {McCoustra}, M.~R.~S., \& {Williams}, D.~A.
  2001, \mnras, 327, 1165

\bibitem[{{Frerking} {et~al.}(1982){Frerking}, {Langer}, \&
  {Wilson}}]{Frerking1982}
{Frerking}, M.~A., {Langer}, W.~D., \& {Wilson}, R.~W. 1982, \apj, 262, 590

\bibitem[{{Hily-Blant} {et~al.}(2010){Hily-Blant}, {Maret}, {Bacmann},
  {Bottinelli}, {Parise}, {Caux}, {Faure}, {Bergin}, {Blake}, {Castets},
  {Ceccarelli}, {Cernicharo}, {Coutens}, {Crimier}, {Demyk}, {Dominik},
  {Gerin}, {Hennebelle}, {Henning}, {Kahane}, {Klotz}, {Melnick}, {Pagani},
  {Schilke}, {Vastel}, {Wakelam}, {Walters}, {Baudry}, {Bell}, {Benedettini},
  {Boogert}, {Cabrit}, {Caselli}, {Codella}, {Comito}, {Encrenaz}, {Falgarone},
  {Fuente}, {Goldsmith}, {Helmich}, {Herbst}, {Jacq}, {Kama}, {Langer},
  {Lefloch}, {Lis}, {Lord}, {Lorenzani}, {Neufeld}, {Nisini}, {Pacheco},
  {Phillips}, {Salez}, {Saraceno}, {Schuster}, {Tielens}, {van der Tak}, {van
  der Wiel}, {Viti}, {Wyrowski}, \& {Yorke}}]{Hily-Blant2010}
{Hily-Blant}, P., {Maret}, S., {Bacmann}, A., {et~al.} 2010, \aap, 521, L52

\bibitem[{{Hogerheijde} \& {van der Tak}(2000)}]{ratran}
{Hogerheijde}, M.~R. \& {van der Tak}, F.~F.~S. 2000, \aap, 362, 697

\bibitem[{{Hollenbach} {et~al.}(2009){Hollenbach}, {Kaufman}, {Bergin}, \&
  {Melnick}}]{Hollenbach2009}
{Hollenbach}, D., {Kaufman}, M.~J., {Bergin}, E.~A., \& {Melnick}, G.~J. 2009,
  \apj, 690, 1497

\bibitem[{{Limbach} {et~al.}(2006){Limbach}, {Buntkowsky}, {Matthes},
  {GrŸndemann}, {Pery}, {Walaszek}, \& {Chaudret}}]{limbach2006}
{Limbach}, H.-H., {Buntkowsky}, G., {Matthes}, J., {et~al.} 2006, ChemPhysChem,
  7, 551

\bibitem[{{Linsky} {et~al.}(2006){Linsky}, {Draine}, {Moos}, {Jenkins}, {Wood},
  {Oliveira}, {Blair}, {Friedman}, {Gry}, {Knauth}, {Kruk}, {Lacour}, {Lehner},
  {Redfield}, {Shull}, {Sonneborn}, \& {Williger}}]{Linsky2006}
{Linsky}, J.~L., {Draine}, B.~T., {Moos}, H.~W., {et~al.} 2006, \apj, 647, 1106

\bibitem[{{M{\"u}ller} {et~al.}(2005){M{\"u}ller}, {Schl{\"o}der}, {Stutzki},
  \& {Winnewisser}}]{muller2005}
{M{\"u}ller}, H.~S.~P., {Schl{\"o}der}, F., {Stutzki}, J., \& {Winnewisser}, G.
  2005, Journal of Molecular Structure, 742, 215

\bibitem[{{Ott}(2010)}]{ott2010}
{Ott}, S. 2010, in Astronomical Society of the Pacific Conference Series, Vol.
  434, Astronomical Data Analysis Software and Systems XIX, ed. Y.~{Mizumoto},
  K.-I. {Morita}, \& M.~{Ohishi}, 139

\bibitem[{{Pardanaud} {et~al.}(2007){Pardanaud}, {Michaut}, {Fillion},
  {Vasserot}, \& {Abouaf-Marguin}}]{Pardanaud2007}
{Pardanaud}, C., {Michaut}, X., {Fillion}, J.-H., {Vasserot}, A.-M., \&
  {Abouaf-Marguin}, L. 2007, in Molecules in Space and Laboratory

\bibitem[{{Parise} {et~al.}(2012){Parise}, {Du}, {Liu}, {Belloche},
  {Wiesemeyer}, {G{\"u}sten}, {Menten}, {H{\"u}bers}, \& {Klein}}]{Parise2012}
{Parise}, B., {Du}, F., {Liu}, F.-C., {et~al.} 2012, \aap, 542, L5

\bibitem[{{Persson} {et~al.}(2013){Persson}, {J{\o}rgensen}, \& {van
  Dishoeck}}]{persson2012}
{Persson}, M.~V., {J{\o}rgensen}, J.~K., \& {van Dishoeck}, E.~F. 2013, \aap,
  549, L3

\bibitem[{{Pilbratt} {et~al.}(2010){Pilbratt}, {Riedinger}, {Passvogel},
  {Crone}, {Doyle}, {Gageur}, {Heras}, {Jewell}, {Metcalfe}, {Ott}, \&
  {Schmidt}}]{pilbratt2010}
{Pilbratt}, G.~L., {Riedinger}, J.~R., {Passvogel}, T., {et~al.} 2010, \aap,
  518, L1

\bibitem[{{Pineda} {et~al.}(2012){Pineda}, {Maury}, {Fuller}, {Testi},
  {Garc{\'{\i}}a-Appadoo}, {Peck}, {Villard}, {Corder}, {van Kempen}, {Turner},
  {Tachihara}, \& {Dent}}]{pineda2012}
{Pineda}, J.~E., {Maury}, A.~J., {Fuller}, G.~A., {et~al.} 2012, \aap, 544, L7

\bibitem[{{Roberts} {et~al.}(2002){Roberts}, {Herbst}, \&
  {Millar}}]{roberts2002}
{Roberts}, H., {Herbst}, E., \& {Millar}, T.~J. 2002, \mnras, 336, 283

\bibitem[{{Roelfsema} {et~al.}(2012){Roelfsema}, {Helmich}, {Teyssier},
  {Ossenkopf}, {Morris}, {Olberg}, {Shipman}, {Risacher}, {Akyilmaz},
  {Assendorp}, {Avruch}, {Beintema}, {Biver}, {Boogert}, {Borys}, {Braine},
  {Caris}, {Caux}, {Cernicharo}, {Coeur-Joly}, {Comito}, {de Lange},
  {Delforge}, {Dieleman}, {Dubbeldam}, {de Graauw}, {Edwards}, {Fich},
  {Flederus}, {Gal}, {di Giorgio}, {Herpin}, {Higgins}, {Hoac}, {Huisman},
  {Jarchow}, {Jellema}, {de Jonge}, {Kester}, {Klein}, {Kooi}, {Kramer},
  {Laauwen}, {Larsson}, {Leinz}, {Lord}, {Lorenzani}, {Luinge}, {Marston},
  {Mart{\'{\i}}n-Pintado}, {McCoey}, {Melchior}, {Michalska}, {Moreno},
  {M{\"u}ller}, {Nowosielski}, {Okada}, {Orlea{\'n}ski}, {Phillips}, {Pearson},
  {Rabois}, {Ravera}, {Rector}, {Rengel}, {Sagawa}, {Salomons},
  {S{\'a}nchez-Su{\'a}rez}, {Schieder}, {Schl{\"o}der}, {Schm{\"u}lling},
  {Soldati}, {Stutzki}, {Thomas}, {Tielens}, {Vastel}, {Wildeman}, {Xie},
  {Xilouris}, {Wafelbakker}, {Whyborn}, {Zaal}, {Bell}, {Bjerkeli}, {De Beck},
  {Cavali{\'e}}, {Crockett}, {Hily-Blant}, {Kama}, {Kaminski}, {Lefl{\'o}ch},
  {Lombaert}, {de Luca}, {Makai}, {Marseille}, {Nagy}, {Pacheco}, {van der
  Wiel}, {Wang}, \& {Y{\i}ld{\i}z}}]{roelfsema2012}
{Roelfsema}, P.~R., {Helmich}, F.~P., {Teyssier}, D., {et~al.} 2012, \aap, 537,
  A17

\bibitem[{{Scribano} {et~al.}(2010){Scribano}, {Faure}, \&
  {Wiesenfeld}}]{Scribano2010}
{Scribano}, Y., {Faure}, A., \& {Wiesenfeld}, L. 2010, \jcp, 133, 231105

\bibitem[{{Smith} {et~al.}(1997){Smith}, {Huang}, \& {Kay}}]{Smith1997}
{Smith}, S.~R., {Huang}, C., \& {Kay}, B.~D. 1997, J. Phys. Chem. B, 101, 6123

\bibitem[{{Sultanov} \& {Balakrishnan}(2005)}]{Sultanov2005}
{Sultanov}, R.~A. \& {Balakrishnan}, N. 2005, \apj, 629, 305

\bibitem[{{Taquet} {et~al.}(2013){Taquet}, {Peters}, {Kahane}, {Ceccarelli},
  {L{\'o}pez-Sepulcre}, {Toubin}, {Duflot}, \& {Wiesenfeld}}]{Taquet2012}
{Taquet}, V., {Peters}, P.~S., {Kahane}, C., {et~al.} 2013, \aap, 550, A127

\bibitem[{{Vastel} {et~al.}(2010){Vastel}, {Ceccarelli}, {Caux}, {Coutens},
  {Cernicharo}, {Bottinelli}, {Demyk}, {Faure}, {Wiesenfeld}, {Scribano},
  {Bacmann}, {Hily-Blant}, {Maret}, {Walters}, {Bergin}, {Blake}, {Castets},
  {Crimier}, {Dominik}, {Encrenaz}, {G{\'e}rin}, {Hennebelle}, {Kahane},
  {Klotz}, {Melnick}, {Pagani}, {Parise}, {Schilke}, {Wakelam}, {Baudry},
  {Bell}, {Benedettini}, {Boogert}, {Cabrit}, {Caselli}, {Codella}, {Comito},
  {Falgarone}, {Fuente}, {Goldsmith}, {Helmich}, {Henning}, {Herbst}, {Jacq},
  {Kama}, {Langer}, {Lefloch}, {Lis}, {Lord}, {Lorenzani}, {Neufeld}, {Nisini},
  {Pacheco}, {Pearson}, {Phillips}, {Salez}, {Saraceno}, {Schuster}, {Tielens},
  {van der Tak}, {van der Wiel}, {Viti}, {Wyrowski}, {Yorke}, {Cais}, {Krieg},
  {Olberg}, \& {Ravera}}]{vastel2010}
{Vastel}, C., {Ceccarelli}, C., {Caux}, E., {et~al.} 2010, \aap, 521, L31

\bibitem[{{Weck} {et~al.}(2006){Weck}, {Balakrishnan}, {Brand\~{a}o}, {Rosa},
  \& {Wang}}]{weck2006}
{Weck}, P.~F., {Balakrishnan}, N., {Brand\~{a}o}, J., {Rosa}, C., \& {Wang}, W.
  2006, \jcp, 124, 074308

\bibitem[{Woon \& Herbst(2009)}]{Woon2009}
Woon, D.~E. \& Herbst, E. 2009, \apjs, 185, 273

\end{thebibliography}

\end{document}